\documentclass[preprint,prb,citeautoscript]{revtex4}
\usepackage{graphicx}
\usepackage{dcolumn}
\usepackage{bm}
\makeatletter
\renewcommand\@biblabel[1]{(#1)}
\makeatother

\begin{document}

\begin{titlepage}

\title{Multiple Localized States and Magnetic Orderings in Partially Open Zigzag Carbon Nanotube Superlattices: An Ab Initio Study}

\author{Bing Huang$^1$'$^2$, Zuanyi Li$^3$, Young-Woo Son$^4$, Gunn Kim$^5$\footnote{gunnkim@khu.ac.kr}, Wenhui Duan$^1$\footnote{dwh@phys.tsinghua.edu.cn}, and Jisoon Ihm$^2$}
\address{$^1$ Department of Physics, Tsinghua University, Beijing
100084, People's Republic of China \\ $^2$ FPRD and Department of
Physics and Astronomy, Seoul National University, Seoul 151-747,
Republic of Korea \\ $^3$ Department of Physics, University of
Illinois at Urbana-Champaign, Urbana, Illinois 61801-3080, USA \\
$^4$ Korea Institute for Advanced Study, Seoul
130-722, Republic of Korea \\
$^5$ Department of Physics, Kyung Hee University, Seoul 130-701,
Korea}
\date{\today}

\begin{abstract}

Using ab initio calculations, we examine the electronic and magnetic
properties of partially open (unzipped) zigzag carbon nanotube (CNT)
superlattices. It is found that depending on their opening degree,
these superlattices can exhibit multiple localized states around the
Fermi energy. More importantly, some electronic states confined in
some parts of the structure even have special magnetic orderings. We
demonstrate that, as a proof of principle, some partially open
zigzag CNT superlattices are by themselves giant (100\%)
magnetoresistive devices. Furthermore, the localized (and
spin-polarized) states as well as the band gaps of the superlattices
could be further modulated by external electric fields perpendicular
to the tube axis, and a bias voltage along the tube axis may be used
to control the conductance of two spin states. We believe that these
results will open the way to the production of novel nano-scale
electronic and spintronic devices.

\end{abstract}


\maketitle

\draft

\vspace{2mm}

\end{titlepage}

\section{Introduction}

Low dimensional sp$^2$-bonded carbon materials have attracted tremendous
attention because of their unique electronic properties and
potential applications in nano-scale devices\cite{Charlier-2007,
Avouris-2007,  Geim-2007, A. H. Castro Neto-2009}. For examples,
depending on their diameters and chiralities, carbon nanotubes
(CNTs) can be either metallic or semiconducting, and the later ones
(\emph{e.g.}, semiconducting zigzag CNTs) are suitable for making
nano-scale electronic devices\cite{Tans-1998, Javey-2003,
Charlier-2007, Avouris-2007}. Graphene nanoribbons (GNRs),
which could be regarded as unrolled CNTs, also exhibit unique electronic structures strongly dependent on the
orientation of their edges and width\cite{Fujita-1996, Nakada-1996,
Wakabayashi-1999, Son-2006}. On the other hand, superlattices as
well as various quantum structures based on CNTs or GNRs have been
the subject of active research in the past decades, and
one-dimensional quantum wells and zero-dimensional quantum dots have
been shown to form by either CNT-based\cite{Chico-1998, Kilic-2000,
Gulseren-2003, Chico-2004, Zhang-2005, GKim-2007} or
GNR-based\cite{Trauzettel-2007, Fernandez-2007, Wang-2008,
Park-2008, Topsakal-2008, Sevincli-2008, Wang-2009, Park-2009,
Guclu-2009} structures.

So far, most studies of CNTs and GNRs have been conducted
independently. The experimental realization of GNRs generally relies
on standard e-beam (or microscope) lithography\cite{Han-2007,
Chen-2007, Tapaszto-2008}, chemical method\cite{Li-2008}, or
synthetic method\cite{Yang-2008}. Quite recently, however, several
promising methods were developed by using CNTs to fabricate GNRs or
CNT-GNR junctions via longitudinal cutting\cite{Cano-Marquez-2009,
Elias-2009, Kosynkin-2009, Zenxing-2009, Jiao-2009, Jiao-2010,
Jiao-2010NR, Terrones-2010} such as chemical attack or plasma
etching to unzip CNTs\cite{Kosynkin-2009, Zenxing-2009, Jiao-2009,
Jiao-2010, Jiao-2010NR}. Besides, the side wall of CNTs can be
opened longitudinally by intercalation of lithium atoms or
transition metal clusters and then ammonia followed by
exfoliation\cite{Cano-Marquez-2009, Elias-2009}. More interestingly,
the degree of stepwise opening in CNTs can be controlled by the
experimental conditions, such as the amount of oxidizing agents,
lithium atoms, or transition metal clusters, and partially open
(unzipped) CNTs (or CNT-GNR junctions) have been observed in
transmission electron microscopy images\cite{Kosynkin-2009,
Zenxing-2009, Cano-Marquez-2009, Elias-2009, Jiao-2010,
Jiao-2010NR}. These experimental progresses imply that CNTs and GNRs
can be combined together with perfect-atomic-interfaces and new
superlattices with CNT-GNR junctions can be
realized\cite{Kosynkin-2009, Zenxing-2009, Cano-Marquez-2009,
Elias-2009, Jiao-2010, Jiao-2010NR}. In such nanostructures, the
openings of CNTs (i.e., curved GNR parts) would offer various edges
which are suitable for doping, adsorption, and chemical
functionalization. In particular, zigzag edges have been found to
exhibit localized edge states and unique magnetic ordering in GNRs
and graphene nanostructures\cite{Son-2006, Bing-2008,
Fernandez-2007, Topsakal-2008, Wang-2008, Wang-2009, Bing-PRL,
Guclu-2009}. Being composed of both a CNT and a curved GNR (with
plenty edge structures), a partially open CNT is therefore expected
to have more diverse electronic and magnetic properties than a
pristine CNT or GNR, and these properties may be sensitively
dependent on the opening degree.

Previous theoretical works only focused on partially open (unzipped)
\emph{armchair} CNTs and demonstrated that they are by themselves
magnetoresistive\cite{Santos-2009} and spin-filter
devices\cite{Bing-2009, Bin-2010}. In contrast, as far as we know,
the physical properties of partially open \emph{zigzag} CNTs have
been poorly investigated until now. In this work, electronic and
magnetic properties of superlattices made of partially open zigzag
CNTs are investigated using spin-polarized density functional theory
(DFT) calculations. Depending on the opening degree, the
superlattices are found to show semiconducting or spin-polarized
semiconducting/metallic behaviors. Besides, some localized states
appear around the Fermi level and some of them could serve as
quantum well states which cannot be achieved in partially open
armchair CNTs. Especially, magnetic (spin) ordering can be realized
in these superlattice structures and mainly depends on the opening
width rather than length unlike partially open armchair CNTs. Our
calculations further suggest some special partially open zigzag CNTs
superlattices can also act as giant magnetoresistive devices under
magnetic fields. Furthermore, the band gap as well as the localized
states could be modulated by external transverse electric fields,
and the conductance of two spin states may be tuned by a bias
voltage along the tube axis.

\section{Computational Methods and Models}

We carried out electronic structure calculations using the projector
augmented wave (PAW) potentials\cite{PAW} and the generalized
gradient approximation (GGA) with the Perdew-Burk-Ernzerhof (PBE)
functional\cite{PBE} implemented in the Vienna \emph{Ab initio}
Simulation Package (VASP)\cite{VASP}. The energy cutoff for a plane
wave basis set is 400 eV. Our models are optimized until energies
are converged to 10$^{-5}$ eV and atomic forces are smaller than
0.02 eV/\AA. The supercell is sufficient large to ensure that the
vacuum space between the adjacent CNTs are at least 10 \AA~ to avoid
the interaction. Three Monkhost-Pack \emph{k}-point meshes are
employed, yielding an $\sim$ 1 meV per atom convergence of the total
energy. The optimized geometrical structures are used to calculate
the electronic and magnetic structures under uniform external
electric field ($E_{\rm ext}$), implemented using a dipole layer in
the vacuum as in the work of Neugebauer and
Scheffler\cite{Neugebauer}.

%fig01
\begin{figure}[tbp]
\includegraphics[width=0.8\textwidth]{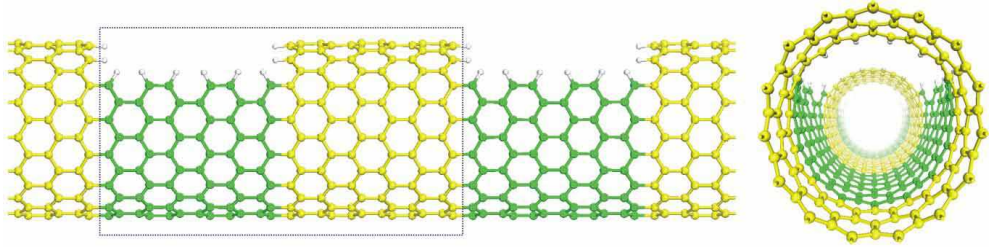}%12.0cm
\caption{Structures of partially open ($n$, $m$) CNTs, NT($n$,
$m$, $L_{t}$)-ONT($W$, $L_{o}$). Yellow and green balls represent
carbon atoms on the perfect CNT and the open CNT parts,
respectively. The openings are terminated with hydrogen atoms,
represented by small white balls. The rectangle marks one
supercell of partially open CNTs. This figure shows the structure
of NT(15, 0, 6)-ONT(4, 6) representing a partially open (15, 0)
CNT which has 6 (6) C-C dimer lines in the perfect CNT part (open
part) along the tube axis, and the missing rows in the opening is
4.}
\end{figure}

Before the discussion of our models and results, we first introduce
the definition of partially open CNT superlattices following the
convention of previous work\cite{Bing-2009}. As shown in Figure 1, a
partially open CNT consists of two parts: a perfect CNT part (yellow
color) and an opening (green color). We represent the system as
NT($n$, $m$, $L_{t}$)-ONT($W$, $L_{o}$) where $n$ and $m$ correspond
to the chiral vector of the CNT, $L_{t}$ ($L_{o}$) is the length in
the tube axis direction of the perfect (open) CNT in the units of
columns of carbon atoms. $W$ is the missing width in units of carbon
rows for the opening. For example, we denote the structure (in
Figure 1) as NT(15, 0, 6)-ONT(4, 6). Such a supercell is repeated
periodically in the tube axis direction to form a superlattice.

\section{Results and discussion}

A zigzag CNT could be either semiconducting or almost metallic at
room temperature\cite{Charlier-2007}. Thus, it is expected that
partially open CNTs obtained from distinct zigzag CNTs may show
different features depending on their chiralities. The (13, 0) CNT
(with a band gap of $\sim$ 0.72 eV in our caculations) and the (15,
0) CNT (with a band gap of $\sim$ 0.04 eV) are considered to
construct superlattices. We find that the electronic properties,
especially the magnetic properties, of both kinds of partially open
zigzag CNT superlattices depend more sensitively on the cutting
width than the cutting length; this is evidently different from the
partially open armchair CNTs\cite{Bing-2009, Santos-2009, Bin-2010}.
Furthermore, for both (13, 0) and (15, 0) CNTs, the
spin-polarization takes place at the discontinuous zigzag edges of
the opening when the cutting width $W >$ 5. Two different cutting
widths ($W = 4$ and $W = 6$) are chosen to represent nonmagnetic
(NM) and spin-polarized system, respectively. In the following, we
keep the total length of the superlattices as 2.56 nm ($L_{t}$ +
$L_{o}$ = 12), and then change the cutting length $L_{o}$. The
results are listed in Table I.

\begin{table*}
\caption{The relation between the energy band gap $E_g$ (eV) and
the cutting length $L_{o}$ of partially open (13, 0) and (15, 0)
CNT superlattices for two different cutting width ($W$) [$L_{t}$ +
$L_{o}$ = 12]. As in  $W = 6$ case, the energy differences $\Delta
E$ (eV/cell) between antiferromagnetic (AFM) ground states and
nonmagnetic (NM) states are also shown.}
\begin{ruledtabular}
\begin{tabular}{lcccclccccc}
\multicolumn{4}{c}{NT(13, 0, $L_{t}$)-ONT ($W$, $L_{o}$)} & &
\multicolumn{4}{c}{NT(15, 0, $L_{t}$)-ONT ($W$, $L_{o}$)} \\
\hline
~~~~~~~~~$E_g$ & $L_{o}$ = 4 & $L_{o}$ = 6 & $L_{o}$ = 8 & &  ~~~~~~~~~$E_g$ & $L_{o}$ = 4 & $L_{o}$ = 6 & $L_{o}$ = 8\\[2pt]
\hline
$W=4$ (NM) &0.330&0.295&0.224 & & $W=4$ (NM) &0.321&0.365&0.384 \\[2pt]
\hline
~~~~~~~~~(AFM) &0.458&0.462&0.475 & &  ~~~~~~~~~(AFM) &0.193&0.189&0.174 \\[2pt]
\raisebox{2.3ex}[0pt]{$W=6$} $\Delta E$ &-0.164&-0.135&-0.064 & & \raisebox{2.3ex}[0pt]{$W=6$} $\Delta E$  &-0.057 &-0.043&-0.020 \\[2pt]
\end{tabular}
\end{ruledtabular}
\end{table*}

For $W=4$, the overall geometric structures change slightly when the
cutting length is short. As the cutting length becomes longer, the
middle part of the opening is widened to release the compressive
stress and lower the total energy, which is similar to partially
open armchair CNTs\cite{Bing-2009, Santos-2009, Bin-2010,
Rangel-2009} and agrees with experimental
observations\cite{Cano-Marquez-2009, Elias-2009, Kosynkin-2009,
Zenxing-2009, Jiao-2010, Jiao-2010NR}. The partially open (13, 0)
and (15, 0) CNT superlattices are still semiconductors without any
magnetic ordering. However, their band gaps change much compared
with perfect (15, 0) and (13, 0) CNTs, as listed in Table I.

%fig02
\begin{figure}[tbp]
\includegraphics[width=0.8\textwidth]{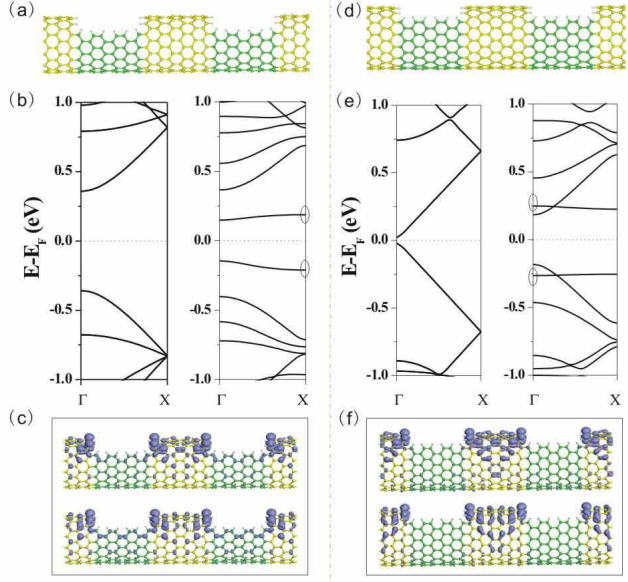}%10.0cm
\caption{(a) The optimized structure of NT(13, 0, 6)-ONT(4, 6)
superlattice (side view). (b) From left to right, the band
structures of the perfect (13, 0) CNT with six times primitive unit
cell and NT(13, 0, 6)-ONT(4, 6) superlattice. (c) Side view of the
charge density on the top valence band (upper figure) and the bottom
conduction band (lower figure) at the X-point of the band structure
in (b)(marked with circles). (d) The optimized structure of NT(15,
0, 6)-ONT(4, 6) superlattice (side view). (e) From left to right,
the band structures of perfect (15, 0) CNT with six times primitive
unit cell and NT(15, 0, 6)-ONT(4, 6) superlattice. (f) Side view of
the charge density on the top valence band (upper figure) and the
bottom conduction band (lower figure) at the $\Gamma$-point of the
band structure in (e)(marked with circles). The Fermi level is set
to zero.}
\end{figure}

As an example, Figure 2 shows the electronic properties of NT(13, 0,
6)-ONT(4, 6) and NT(15, 0, 6)-ONT(4, 6) superlattices. The band
structures of perfect (13, 0) and (15, 0) CNTs, which are calculated
in the supercell six times than the primitive unit cell, are also
displayed in the left panels of Figs. 2b and 2e for comparison. The
bands around the Fermi level of perfect (13, 0) and (15, 0) CNTs are
doubly-degenerate, but the degeneracy disappears due to cylindrical
symmetry breaking. This is evident from the increased numbers of
visible bands in the right panels of Figs. 2b and 2e. Two bands with
small dispersions arise around the Fermi level in the NT(13, 0,
6)-ONT(4, 6) superlattice, and consequently, the band gap of NT(13,
0, 6)-ONT(4, 6) is $\sim$ 50\% smaller than that of the perfect one.
Charge density analysis (Figure 2c) shows that most of charge
densities of the two bands are located at the top of the perfect CNT
part (especially located at the zigzag edges) and contributed mainly
by the \emph{p$_{z}$} orbitals of carbon atoms. The physical origin
of these localized states is similar to that of zigzag graphene
nanoribbons (ZGNRs). Previous works show that the zigzag edges of
graphene (or ZGNRs) induce localized edge states around the Fermi
level\cite{Fujita-1996, Nakada-1996, Wakabayashi-1999, Bing-2008,
Zuanyi-PRL}. As the cutting length $L_{o}$ increases from 4 to 8,
the energy differences between the two localized states
(\emph{i.e.}, band gap between the top valence and bottom conduction
bands) decrease from 0.330 eV to 0.224 eV (Table I), and the two
bands become a little more dispersive due to the hybridization with
$\pi$ and $\pi^*$ bands (the second highest valence band and second
lowest conduction band) near the $\Gamma$ point. Similar to NT(13,
0, 6)-ONT(4, 6) superlattice, there are also two new flat bands
located close to the top valence band and the bottom conduction band
for NT(15, 0, 6)-ONT(4, 6) superlattice. The two flat bands are also
mainly located at the zigzag edges of the top perfect CNT part, as
shown in Figure 2f. The origin of these localized states is the same
as in NT(13, 0, 6)-ONT(4, 6) superlattice. When the cutting length
$L_{o}$ increases from 4 to 8, the band gap of partially open (15,
0) CNT superlattices increases slightly from 0.32 eV to 0.38 eV,
whereas the energy difference between the two localized states
increases from 0.40 eV to 0.63 eV.

%fig03
\begin{figure}[tbp]
\includegraphics[width=0.8\textwidth]{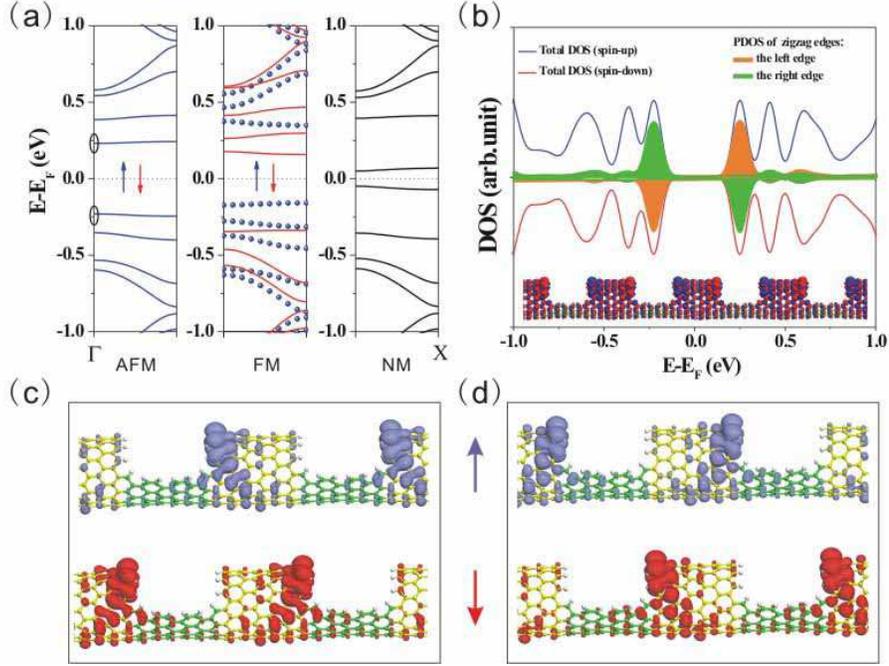}%12.0cm
\caption{(a) From left to right, the band structures of NT(13, 0,
6)-ONT(6, 6) superlattice for AFM, FM, and NM states, respectively.
(b) The total density of states (DOS) of NT(13, 0, 6)-ONT(6, 6)
superlattice in the AFM ground state. The partial density of states
(PDOS) of two zigzag edges at the opening is also plotted as green
(for right) and orange (for left) filled area under DOS curve. The
spatial spin density distribution of AFM ground state is shown as
inset. The Fermi level is set to zero. (c) The spin-up and -down
charge densities (side views) of the top valence band at the
$\Gamma$-point in the AFM state [the lower circle in (a)]. (d) The
spin-up and -down charge densities (side views) of the bottom
conduction band at the $\Gamma$-point in the AFM state [the upper
circle in (a)]. In these figures, blue and red colors represent the
spin-up and spin-down states, respectively.}
\end{figure}

The situation is much more interesting for $W$ = 6. The ground state
of these superlattices converts from NM to spin-polarized. There are
two stable spin-polarized states, called the AFM (ground) and FM
states, similar to the case of ZGNRs\cite{Son-2006, Bing-2008,
Fujita-1996, Son-Nature}. The AFM (FM) configuration exhibits
antiferromagnetic (ferromagnetic) coupling between two zigzag edges
in one opening and the ferromagnetic coupling at each zigzag edge,
as shown in the inset of Figures 3b or Figure 4b. For $W$ = 6, the
band gap changes little with different cutting length, as listed in
Table I; this is evidently different from the case of $W$ = 4. It
means that the energy gap opening mechanism of the spin-polarized
case is different from that of the NM case, as explained in the
following. The energy difference between the AFM ground states and
NM states, however, decreases much as cutting length $L_{o}$
increases.

Figures 3 and 4 show the electronic and magnetic properties of
NT(13, 0, 6)-ONT(6, 6) and NT(15, 0, 6)-ONT(6, 6) superlattices,
respectively. NT(13, 0, 6)-ONT(6, 6) superlattice has four flat
bands around the Fermi level for both spin-polarized (AFM and FM)
states and NM state, as shown in Figure 3a. Moreover, the top
valence and bottom conduction bands of the AFM configurations are
mostly confined at the zigzag edges of the opening, and behave as
spin-polarized edge states, as shown in the partial density of
states (PDOS) and the charge density in Figures 3b-3d. Comparison
with the case of $W =4$ indicates that long enough topological
zigzag edges induce spin instability toward spin-band splitting,
which agrees with previous finding on ZGNRs and CNTs\cite{Son-2006,
Bing-2008, Son-Nature, Bing-2009, Santos-2009, Bin-2010}. Another
two flat states of each spin (the second highest valence band and
the second lowest conduction band) are mainly located at the bottom
of the tube, which behave as quantum well states (data not shown).
As shown in Figures 3c and 3d, the oppositely oriented spin states
are mostly located at the opposite (left and right) sides of open
zigzag edges. Therefore, if an additional electron (hole) is
introduced in this system, it will be confined at one zigzag edge of
the opening with a certain spin direction. The flat bands around the
Fermi level in the FM state ($\sim$ 0.025 eV/cell higher than the
AFM state) have similar properties to the corresponding flat bands
of the AFM state. It is also interesting that as the cutting length
increases from 4 to 8, the dispersions and energies of the two flat
bands associated with localized states change little but the
energies of the other two flat bands associated with quantum well
states change much (data not shown). Therefore, the existence of
spin-polarized localized states and the band gaps of AFM ground
state are insensitive to the cutting length but sensitive to the
cutting width.

%fig04
\begin{figure}[tbp]
\includegraphics[width=0.8\textwidth]{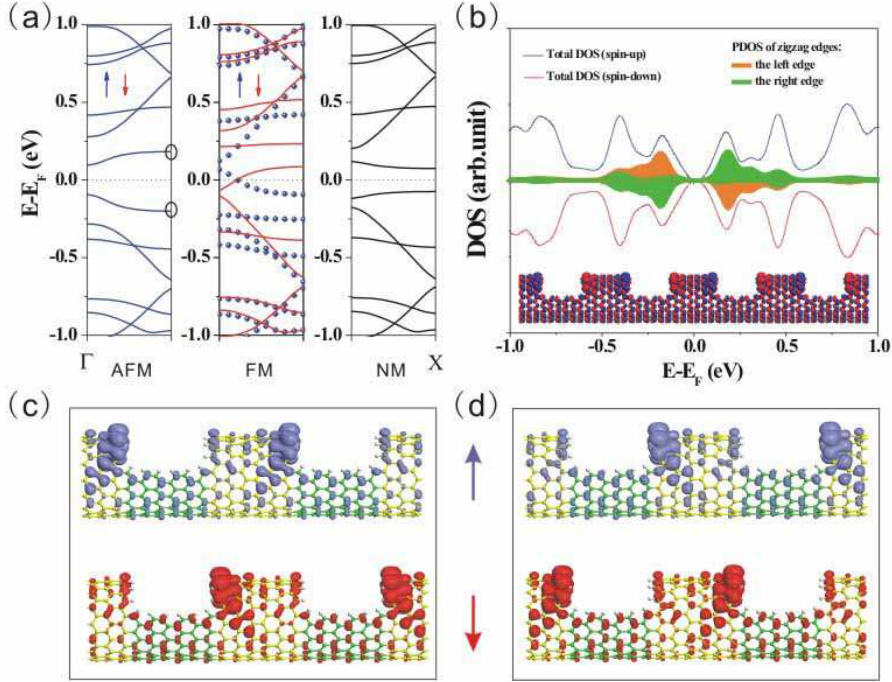}%12.0cm
\caption{(a) From left to right, the band structures of NT(15, 0,
6)-ONT(6, 6) superlattice for AFM, FM, and NM states, respectively.
(b) The total DOS of NT(15, 0, 6)-ONT(6, 6) superlattice in the AFM
ground state. The PDOS of two zigzag edges at opening is also
plotted as green (for right) and orange (for left) filled area under
DOS curve. The spatial spin density distribution of the AFM ground
state is shown as inset. The Fermi level is set to zero. (c) The
spin-up and -down charge densities (side views) of the top valence
band at the X-point in the AFM state [the lower circle in (a)]. (d)
The spin-up and -down charge densities (side views) of the bottom
conduction band at the X-point in the AFM state [the upper circle in
(a)]. In these figures, blue and red colors represent the spin-up
and spin-down states, respectively.}
\end{figure}

As shown in Figure 4, NT(15, 0, 6)-ONT(6, 6) superlattice exhibits
an AFM ground state with 0.189 eV band gap, similar to NT(13, 0,
6)-ONT(6, 6). The top valence band and the bottom conduction band of
the AFM state have a small dispersion with a flattened tail near the
X-point (Figure 4a) and the charge density of the two bands at the
X-point are located at the opposite (left or right) zigzag edge of
the opening (Figure 4c and 4d). At the same time, the zigzag edge
states are hybridized with dispersive $\pi$ and $\pi^*$ bands near
the $\Gamma$ point, as shown in Figure 4a and 4b, which is obviously
different from the above NT(13, 0, 6)-ONT(6, 6) case. Unexpectedly,
the FM state ($\sim$ 0.030 eV/cell higher than the AFM state) is
metallic with $\sim 1.0 \mu_{\rm B}$ moment. As the cutting length
increases from 4 to 8, the band gap of the AFM state decreases by
0.02 eV (Table I), and the FM is always metallic with $\sim 1.0
\mu_{\rm B}$ moment, independent on the cutting length. Similar
results are obtained in our calculations for NT(12, 0, 6)-ONT(6, 6)
superlattice, indicating that the FM states of partially open CNT
superlattices made of metallic zigzag CNTs are metallic when the
cutting width is large enough ($W$ $>$ 5).

In addition, we keep the opening degree ($L_{o}$ = 6, $W$ = 6) of
the CNT superlattices unchanged, and then increase the total length
($L_{t}$ increases as well) to study the dependence of electronic
properties (especially the energy of the localized states) on
$L_{t}$. The results are summarized in Table II. For NT(13, 0,
$L_{t}$)-ONT(6, 6) and NT(15, 0, $L_{t}$)-ONT(6, 6) superlattices,
clearly, the electronic properties (band gaps) are insensitive to
the length $L_{t}$, and the overall band structure around the Fermi
level retains the same for different $L_{t}$ (not shown). The energy
differences between the AFM state and NM states increase slightly,
indicating the AFM ground state becomes a little more stable for
longer $L_{t}$.

\begin{table}
\caption{The band gap $E_g$ (eV) and the energy differences
[$\Delta E$ (eV/cell), between AFM ground states and NM states] of
NT(13, 0, $L_{t}$)-ONT(6, 6) and NT(15, 0, $L_{t}$)-ONT(6, 6)
superlattices with different $L_{t}$.}
\begin{ruledtabular}
\begin{tabular}{ccccccccccc}
\multicolumn{4}{c}{NT(13, 0, $L_{t}$)-ONT (6, 6)} & &
\multicolumn{4}{c}{NT(15, 0, $L_{t}$)-ONT (6, 6)} \\
\hline
$L_{t}$ & 6 & 8 & 10 & & $L_{t}$ & 6 & 8 & 10\\[2pt]
\hline
$E_g$ (AFM) &0.462&0.462&0.462 & &  $E_g$ (AFM) &0.189&0.175&0.160 \\[2pt]
$\Delta E$ &-0.135&-0.175&-0.180 & & $\Delta E$ &-0.043 &-0.055&-0.057 \\[2pt]
\end{tabular}
\end{ruledtabular}
\end{table}

Magnetic field $B$ can be applied to change the magnetic (spin)
ordering of ZGNRs and partially unzipped armchair CNTs from the AFM
configuration to FM configuration\cite{Woo-2008, Palacios-2009,
Santos-2009}. It has been estimated that the switching $B$ can be as
low as 0.03 T at the liquid Helium temperature (4K) for
ZGNRs\cite{Palacios-2009}. Similarly, it is expected that a magnetic
field can also be used to turn the spin ordering of partially open
zigzag CNT superlattices from AFM (ground state) configuration to FM
configuration. Based on the unique electronic properties of
partially open (15, 0) CNT superlattices, we propose a
magnetoresistance (MR) device [a finite partially open (15, 0) CNT
superlattice connected to two semi-infinite intact metallic (15, 0)
CNTs], with a very large value of MR. The AFM ground state of
partially open (15, 0) CNT superlattices (with $W$ = 6) is
semiconducting with a band gap of $\sim$ 0.20 eV (Figure 4a), so the
conductance of the AFM state around the Fermi level is zero (for an
infinite superlattice) or near zero (for finite superlattice due to
the coupling of the superlattice to electrodes). A sufficiently
strong magnetic field $B$ will make the superlattice ferromagnetic
and hence metallic (Figure 4a), which leads to new bands (transport
channels) at low energies and a finite conductance near the Fermi
level. Therefore, magnetic field $B$ can produce a dramatic change
in the conductance of the system, which is quantified by the MR,
defined as the relative change of the resistance when a magnetic
field is applied. Based on a conventional definition of MR
(MR$\equiv\frac{{R_{AF} - R_{FM} }}{{R_{AF}  + R_{FM} }} =
\frac{{G_{FM} - G_{AF} }}{{G_{FM} + G_{AF} }} \times 100$), we
expect a MR $\sim$ 100\% can be reached in $\sim$ 0.20 eV energy
range near the Fermi level at low magnetic field $B$ around the
liquid Helium temperature.

%fig05
\begin{figure}[tbp]
\includegraphics[width=0.8\textwidth]{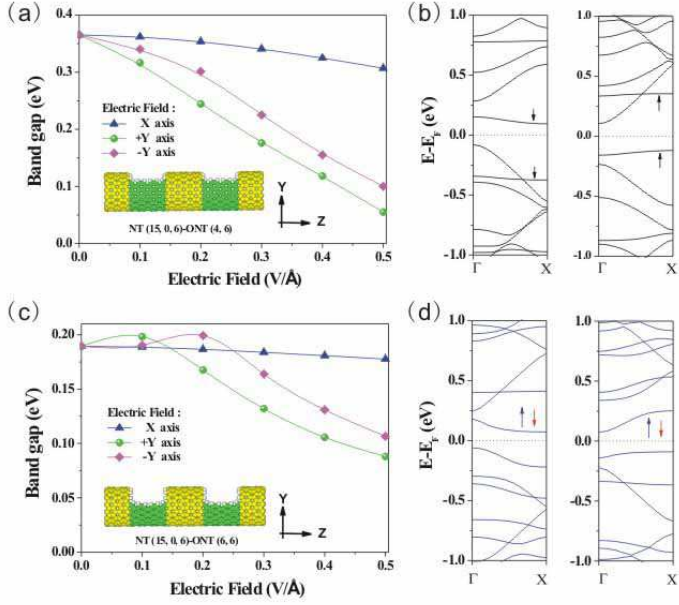}%12.0cm
\caption{Electronic properties of partially open zigzag CNT
superlattices in $E_{\rm ext}$. (a) The band gap as a function of
$E_{\rm ext}$ for NT(15, 0, 6)-ONT(4, 6) superlattice. (b) From left
to right, the band structures of NT(15, 0, 6)-ONT(4, 6) with $E_{\rm
ext}$ = 0.3 V/\AA~ in the $+y$ and $-y$ directions, respectively.
(c) The band gap as a function of $E_{\rm ext}$ for NT(15, 0,
6)-ONT(6, 6) superlattice. (d) From left to right, the band
structures of NT(15, 0, 6)-ONT(6, 6) with $E_{\rm ext}$ = 0.3 V/\AA~
in $+y$ and $-y$ directions, respectively. The inset of (a) and (b)
are the corresponding geometric structures. The Fermi level is set
to zero.}
\end{figure}

The response of CNTs to an external transverse electric field
($E_{\rm ext}$) is also of interest for its future
applications\cite{Sahoo-2005, Fedorov-2005, Son-2005, Son-2007,
Bing-2009}. Our previous works show that the partially open armchair
CNT behaves as an electric switching or a spin-filter under an
$E_{\rm ext}$\cite{Bing-2009, Bin-2010}. So it is interesting to
investigate the properties of partially open zigzag CNTs under
$E_{\rm ext}$. As an example, we have considered NT(15, 0, 6)-ONT(4,
6) and NT(15, 0, 6)-ONT(6, 6) superlattices. For convenience, all
the openings in the superlattices are put in the $+y$ direction,
shown as the insets of Figures 5a and 5b. The $E_{\rm ext}$ has been
applied in three directions ($x$, $+y$, $-y$) perpendicular to the
tube axis ($z$). Figure 5 briefly shows the results of the two
different superlattices under $E_{\rm ext}$.

For NT(15, 0, 6)-ONT(4, 6) superlattice, if $E_{\rm ext}$ is in the
$x$ direction, the band gap changes little ($< 0.06$ eV) even under
0.5 V/\AA~ (Figure 5a). But the situation is opposite if the $E_{\rm
ext}$ is in the $+y$ (or $-y$) direction: the band gap decreases by
$\sim$ 86 \% (or $\sim$ 72 \%). Such changes in the band gap is the
response of localized states to external electric fields. From the
band structures of NT(15, 0, 6)-ONT(4, 6) under $E_{\rm ext}$ = 0.3
V/\AA~ in $+y$ and $-y$ directions (Figure 5b), we can clearly see
that the gap change arises from the upwards/downwards shift of the
two flat bands (associated with the localized states) relative to
the unperturbed bands (Figure 2e) when $E_{\rm ext}$ is applied in
the $-y$/$+y$ direction. Since the two localized states are mainly
composed of the \emph{p$_{z}$} orbital of carbon atoms on the top
part of perfect CNT, their distributions are not uniform in the $y$
direction. Thus, $E_{\rm ext}$ in the $y$ direction can effectively
change the electrostatic potential of the localized states, and
hence their energies. In contrast, there is no obvious difference in
the $x$ direction for these localized states (Figure 2f), so $E_{\rm
ext}$ will not result in a net energy change of each localized
state, and consequently the energy gap does not change dramatically.
Our finding indicates that the energy of localized \emph{p$_{z}$}
orbital states could be strongly influenced by the direction of
charge polarization (dipole direction) induced by $E_{\rm
ext}$\cite{Bing-2009, Tien-2005}. Similar changes of energy gaps are
also observed in the spin-polarized NT(15, 0, 6)-ONT(6, 6)
superlattice (data shown in Figures 5c and 5d) and NT(13, 0,
6)-ONT(6, 6) superlattice (data not shown here). The above
discussion demonstrates that the application of $E_{\rm ext}$ is an
effective way to control the band gap of the superlattices.
Especially, the energy of the localized states can be modulated by
this way for various practice purposes.

Besides the $x$ and $\pm y$ directions, the electric field can
also be applied along the tube axis ($z$ direction). Our previous
works show that if the electric field is transversely applied
across the ZGNRs or partially open armchair CNTs, the occupied and
unoccupied edge (localized) states associated with one spin
orientation close their gap, whereas those associated with the
other spin orientation widen theirs\cite{Son-Nature, Bing-2009,
Zuanyi-JNN}. Then, the system can be forced into a half-metallic
state by an appropriate applied electric field, resulting in an
insulating behavior for one spin and metallic behavior for the
other, and hence the conductance of the system becomes
spin-polarized. The same idea may be applied to the partially open
(15, 0) CNTs in the AFM ground state: with a bias voltage
(electric field) along the tube axis (across the zigzag edges of
the opening), the conductance of the two different spin states
will be changed in different ways, even half-metallicity may be
realized at high bias voltage.

\section{Summary}

In conclusion, our results obtained from ab initio
calculations demonstrate that the superlattices based on partially
open zigzag carbon nanotubes have interesting electronic and
magnetic properties. The localized states can be formed around the
top of perfect CNT parts in such superlattices and the energy gap
can be modified by the size of the opening. The width of the opening
determines whether the ground state of the system is spin-polarized
or not. Interestingly, for the spin-polarized case, the spin states
with antiferromagnetic or ferromagnetic ordering can be confined in
a certain part of these structures. We also find that some partially
open zigzag CNT superlattices are by themselves giant
magnetoresistive devices. Moreover, the band gap as well as the
energy of localized states of the superlattices could be tuned by
external transverse electric fields, and bias voltages along the
tube axis direction may be used to control the conductance of two
different spin states. These properties may be useful for future
device application in nanoelectronics.

\section{Acknowledgments}

We gratefully acknowledge the supports by the KOSEF grant funded by
MEST (Center for Nanotubes and Nanostructured Composites), the
Korean Government MOEHRD, Basic Research Fund No.
KRF-2006-341-C000015, the Core Competence Enhancement Program
(2E2140) of KIST through the Hybrid Computational Science Laboratory
(J.I.), a finacial support (Grant No. KHU-20100119) from Kyung Hee
University (G.K.), and KOSEF grant (Quantum Metamaterials research
center, No. R11-2008-053-01002-0) funded by the MEST (Y.-W.S.). This
work at China was supported by the Ministry of Science and
Technology of China (Grant Nos. 2006CB605105, 2006CB0L0601), and the
National Natural Science Foundation of China (W.D, B.H., and Z.L.).
B.H. also acknowledge the supports by the A3 Foresight Program of
KOSEF-NSFC-JSPS. The computations were performed through the support
of KISTI in Korea.

\newpage

\end{document}